\documentclass[letterpaper]{JHEP3}
\usepackage{epsfig,amsmath,xspace}

\newcommand{\pd}[2]{\frac{\partial #1}{\partial #2}}
\DeclareMathOperator{\tr}{tr}

\newcommand{\G}{\mathcal{G}}
\newcommand{\Sb}{\mathbf{S}}

\title{A note on tachyon actions in string theory}

\author{
Matthew Headrick \\ 
Martin Fisher School of Physics, Brandeis University, Waltham MA 02454, USA \\ 
\email{mph@brandeis.edu}
}

\abstract{
A number of spacetime fields in string theory (notably the metric, dilaton, bosonic and type 0 bulk closed-string tachyon, and bosonic open-string tachyon) have the following property: whenever the spacetime field configuration factorizes in an appropriate sense, the matter sector of the world-sheet theory factorizes into a tensor product of two decoupled theories. Since the beta functions for such a product theory necessarily also factorize, this property strongly constrains the form of the spacetime action encoding those beta functions. We show that this constraint alone---without needing actually to compute any of the beta functions---is sufficient to fix the form of the two-derivative action for the metric-dilaton system, as well as the potential for the bosonic open-string tachyon. We also show that no action consistent with this constraint exists for the closed-string tachyon coupled to the metric and dilaton.
}

\preprint{BRX-TH-602}

\begin{document}

\section{Introduction}

The interplay between world-sheet and spacetime physics is a hallmark of string theory. Each possible spacetime background is represented on the world-sheet as a two-dimensional quantum field theory, and the consistency requirements for world-sheet physics, such as conformal invariance, impose strict constraints on what spacetime backgrounds are (classically) allowed. In this sense the spacetime \emph{theory} is a meta-theory, since it describes all allowed world-sheet theories within some given class. A natural question is then, What constraints does this peculiar theory-within-a-theory structure impose on the spacetime theory? Presumably not every spacetime theory that one can imagine is derivable from string theory.

In this paper we will directly attack that question, working at the classical level. We will concern ourselves with certain string fields possessing a simplifying property we will refer to as \emph{world-sheet factorization}, by which we mean that, whenever the spacetime field configuration factorizes (in an appropriate sense), the world-sheet theory factorizes into a tensor product of two decoupled theories. (Here we are referring to the matter sector of the world-sheet theory, after gauge-fixing.) Suppose, for example, that in the bosonic string theory the spacetime is a product manifold $M_{(1)}\times M_{(2)}$, on which the metric and dilaton factorize as follows:
\begin{align}
ds^2(x^1,x^2) &= ds^2_{(1)}(x^1) + ds^2_{(2)}(x^2)\,, \label{Gfactorize} \\
\Phi(x^1,x^2) &= \Phi_{(1)}(x^1) + \Phi_{(2)}(x^2)\,,
\end{align}
where $x^1$ and $x^2$ represent sets of coordinates on $M_{(1)}$ and $M_{(2)}$ respectively. Then the world-sheet action\footnote{Throughout this paper we will use $S$ for world-sheet actions and $\Sb$ for spacetime actions.}
\begin{equation}\label{WSaction}
S^{ds^2,\Phi}[X] = 
\frac1{2\pi\alpha'}\int d^2\!z\left(
G_{\mu\nu}(X)\partial X^\mu\bar\partial X^\nu + 2\alpha'R_{(2)}\Phi(X)
\right),
\end{equation}
will clearly decompose into
\begin{equation}\label{splits}
S^{ds^2,\Phi}[X^1,X^2] = 
S^{ds^2_{(1)},\Phi_{(1)}}[X^1] +
S^{ds^2_{(2)},\Phi_{(2)}}[X^2]\,,
\end{equation}
and the world-sheet theory will decompose into a tensor product of decoupled theories. The same, of course, applies to the supersymmetric sigma model. Other examples of string fields obeying world-sheet factorization are the $B$-field, the (bosonic and type 0) bulk closed-string tachyon, the open-string gauge field, and the bosonic open-string tachyon. Examples of string fields that do \emph{not} obey world-sheet factorization include compactification moduli, localized closed-string tachyons, transverse scalars on D-branes, and open-string tachyons in superstring theories. In Section 2 below we will discuss the issue of world-sheet factorization in more detail, and explain why it fails in each of these cases. For now, let us proceed to investigate its consequences.

The classical equations of motion for the string fields in spacetime are the requirement that the beta functions of the world-sheet theory vanish. If the world-sheet theory factorizes into two decoupled theories, as in \eqref{splits}, then it is obvious that the beta functions for each factor theory are independent of the other factor theory. In the sigma-model case, we must have
\begin{gather}
\beta^G_{11}[ds^2,\Phi] = \beta^G[ds_{(1)}^2,\Phi_{(1)}]\,, \qquad
\beta^G_{22}[ds^2,\Phi] = \beta^G[ds_{(2)}^2,\Phi_{(2)}]\,, \qquad
\beta^G_{12}[ds^2,\Phi] = 0\,,\label{betassplit1}\\
\beta^\Phi[ds^2,\Phi] = \beta^\Phi[ds_{(1)}^2,\Phi_{(1)}] + \beta^\Phi[ds_{(2)}^2,\Phi_{(2)}]\,.\label{betassplit2}
\end{gather}
(Here in $\beta^\Phi$ we are only including the contribution from the sets of fields $X^1$ and $X^2$, not for example from the ghosts, whereas it is the total dilaton beta function that must vanish to satisfy the spacetime eqations of motion.) It is easy to see that \eqref{betassplit1} and \eqref{betassplit2} are indeed obeyed by the one-loop (i.e.\ two-spacetime-derivative) approximation to the sigma-model beta functions \cite{Callan:1985ia}:
\begin{align}
\label{betaG}
\frac1{\alpha'}\beta^G_{\mu\nu} &=
R_{\mu\nu} + 2\nabla_\mu\partial_\nu\Phi + O(\alpha') \\
\label{betaPhi}
\frac1{\alpha'}\beta^\Phi &= \frac{D}{6\alpha'} -\frac12\nabla^2\Phi + \partial_\omega\Phi\partial^\omega\Phi + O(\alpha')\,.
\end{align}
In fact, the properties \eqref{betassplit1} and \eqref{betassplit2} severely restrict the possible terms that may appear in the beta functions. At the two-derivative level, for example, world-sheet factorization forbids the terms $RG_{\mu\nu}$ and $\partial_\mu\Phi\partial_\nu\Phi$ from appearing in $\beta^G_{\mu\nu}$, which would otherwise be allowed by the diffeomorphism and dilaton shift symmetries.

It should be noted that, although essentially trivial from the world-sheet point of view, from the spacetime point of view the property of factorization is a highly unusual one for a set of field theory equations of motion to possess. Consider, for example, a system as simple as a massless scalar field with a quartic self-coupling; the equation of motion is
\begin{equation}\label{phiEOM}
\nabla^2\phi - \lambda \phi^3 = 0\,.
\end{equation}
If $\phi$ is a sum of a function of $x^1$ and a function of $x^2$,
\begin{equation}\label{phifactorize}
\phi(x^1,x^2) = \phi_{(1)}(x^1) + \phi_{(2)}(x^2)\,,
\end{equation}
then the condition for $\phi(x^1,x^2)$ to solve \eqref{phiEOM} is quite different from the conditions for $\phi_{(1)}$ and $\phi_{(2)}$ separately to solve the same equation. Another simple example is gravity with a minimally coupled massless scalar field, whose equations of motion are
\begin{equation}
\nabla^2\phi = 0\,, \qquad R_{\mu\nu} - \partial_\mu\phi\partial_\nu\phi = 0\,.
\end{equation}
Under the ansatz \eqref{Gfactorize}, \eqref{phifactorize}, the first equation splits nicely. In the second equation, however, the second term contains mixed components $\partial_1\phi_{(1)}\partial_2\phi_{(2)}$, spoiling the factorization.

The approximate sigma-model beta functions \eqref{betaG}, \eqref{betaPhi} possess another property that is rather less obvious from the world-sheet point of view: they are derivable from a spacetime action. Specifically, if we write the functional
\begin{equation}\label{NSaction}
\Sb = 
\alpha'\int d^D\!x\,G^{1/2}e^{-2\Phi}
\left(-\frac{2D}{3\alpha'}+R+4\partial_\mu\Phi\partial^\mu\Phi
+ O(\alpha')
\right),
\end{equation}
then $\beta^G$ and $\beta^\Phi$ are given by linear combinations of the variational derivatives $\delta \Sb/\delta G_{\mu\nu}$ and $\delta \Sb/\delta\Phi$, so that the zeroes of the beta functions coincide with the stationary points of $\Sb$ \cite{Callan:1985ia}.\footnote{More precisely, since the spacetime equations of motion demand that the \emph{total} dilaton beta function vanish, the total spacetime action includes contributions from all sectors of the world-sheet theory, including the ghosts.} Tseytlin has given arguments that this should be true to all orders in $\alpha'$; more specifically, to all orders the action should take the form $\int G^{1/2}e^{-2\Phi}(G^{\mu\nu}\beta^G_{\mu\nu}-4\beta^\Phi)$ \cite{Tseytlin:1988rr}. Certainly on general grounds the existence of a spacetime action would appear to be necessary to guarantee the consistency of classical string theory. For those open-string fields that represent renormalizable boundary couplings, the existence of a spacetime action whose variational derivatives yield the beta functions has been proven; specifically, the action is given by the boundary entropy \cite{Kutasov:2000qp,Friedan:2003yc}.

Our strategy in this paper will be to assume the existence of a spacetime action and derive the consequences of world-sheet factorization for it. (We will also assume basic properties such as diffeomorphism invariance and the masslessness of the dilaton.) For most of the paper, we will work at the two-derivative level. Our main results are:
\begin{enumerate}
\item World-sheet factorization is sufficient to fix the form of the action \eqref{NSaction} for the metric and dilaton. (The $B$-field may also easily be incorporated into the analysis.)
\item World-sheet factorization is sufficient to fix the form of the bosonic open-string tachyon potential; the result agrees with that obtained in Boundary String Field Theory (BSFT) \cite{Gerasimov:2000zp,Kutasov:2000qp}.
\item There does not exist an action for the bulk closed-string tachyon, coupled to the metric and dilaton, obeying world-sheet factorization.
\end{enumerate}
The derivations mentioned in points (1) and (2) are markedly simpler than the standard ones relying in the case of point (1) upon an actual calculation of the world-sheet beta functions, and in the case of point (2) upon the machinery of BSFT (see \cite{Martinec:2002tz,Sen:2004nf} for reviews). We should mention, however, that our derivations leave certain constants in the actions unfixed, such as the height of the open-string tachyon potential (necessary for confirming the Sen conjecture). 

The no-go theorem of point (3) presents a serious problem for the various effective actions proposed recently for the metric-dilaton-tachyon system \cite{Yang:2005rw,Yang:2005rx,Genenberg:2006de,Bergman:2006pd,Suyama:2006wx,Suyama:2007vh,Brandenberger:2007xu,Swanson:2008dt}.\footnote{Aspects of the action for the closed-string tachyon action have been discussed in many other papers, including \cite{Banks:1991sg,Hellerman:2006nx,Moeller:2007mu}.} However, it does not follow from this theorem that there does not exist \emph{any} spacetime action for the closed-string tachyon. Various ways around the theorem are discussed in Section 3, the most likely being that any such action necessarily involves the massive string fields in an essential way. From one point of view this is not surprising: once we include the tachyon in the dynamics, there is no separation of energy scales to prevent the massive string fields from being excited as well. From another point of view (specifically, the world-sheet point of view), however, there is a fundamental difference between the tachyon and the massive string fields: the former represents a renormalizable world-sheet coupling, just like the massless fields, while the latter represent non-renormalizable couplings. By definition, the renormalizable couplings form a closed system under RG flow, so it makes sense to ask whether their beta functions are derivable from an action without including any other string fields. Indeed, this fact is the basis for BSFT, which represents a truncation of the dynamics to the massless and tachyonic open string fields. Although it is well known that the particulars of BSFT do not straightforwardly generalize to the closed-string fields (see \cite{Martinec:2002tz} for an explanation), one might hope that there exists some other way to construct a closed-string field theory just for the tachyonic and massless fields. The no-go theorem presented in this paper would seem to present a serious obstacle for any such theory.

The property of world-sheet factorization has been employed previously, within the context of other investigations, to constrain spacetime actions in string theory (see for example \cite{Myers:1987qx,Cecotti:1990wz,Suyama:2007vh}). Here the focus is on factorization itself, seeing how far it---almost alone---can take us.

The structure of the paper is as follows. In Section 2 we discuss the property of world-sheet factorization in more detail, giving examples of string fields that do and do not obey it. We also remind the reader what it means technically for the beta functions to be ``derivable" from a spacetime action, and spell out some of our technical assumptions.

In Section 3 we consider the implications of world-sheet factorization for various sets of spacetime fields, restricting ourselves to two-derivative actions. We derive the BSFT potential for the bosonic open-string tachyon, as well as the action for that field coupled to the open-string gauge field. We then include gravity, and derive the action for the metric-dilaton system. We also show that no action satisfying world-sheet factorization exists for gravity coupled to more than one scalar field, thereby ruling out an action for the bulk closed-string tachyon coupled to the metric and dilaton.

Finally, in Section 4, we go beyond the two-derivative approximation. In particular, we consider actions that contain arbitrary numbers of derivatives, but only in the form of powers of first derivatives (as in the Nambu-Goto and Born-Infeld actions). For a single scalar, we find that the only such action obeying factorization is of the form $\int e^{\pm(\partial_\mu\phi)^2}$. Since this does not reduce in the limit of small $(\partial_\mu\phi)^2$ to the two-derivative open-string tachyon action (in particular it has no potential term), we learn that the higher-derivative corrections to the latter action either involve more than one derivative acting on the tachyon field, or else involve the other string fields. Finally, we consider an abelian gauge field and find that the most general action respecting factorization is of the form $\int \tr f(F)$ or $\int \det f(F)$, where the arbitrary function $f$ acts on the field strength ${F^\mu}_\nu$ in the usual way that a function acts on a matrix. The Born-Infeld action is a special case of $\int \det f(F)$ with $f(x) = (1+x)^{1/2}$.

Some possible questions for future exploration are:
\begin{enumerate}
\item Can we find a two-derivative action, obeying factorization, for the closed-string tachyon coupled to the metric, dilaton, and massive string fields?
\item In this paper we consider only abelian gauge fields. How does world-sheet factorization constrain non-abelian generalizations of the Born-Infeld action?
\item How does world-sheet factorization constrain higher-derivative corrections to the action for the metric-dilaton system?
\item Finally, what about string fields that do not obey world-sheet factorization? Is there some principle, reflecting their world-sheet origin, that constrains their spacetime actions in the way that factorization does for the fields considered in this paper?
\end{enumerate}

\section{General principles}

\subsection{World-sheet factorization}

As stated in the introduction, we apply the term ``world-sheet factorization" to any string field for which the world-sheet theory factorizes into a tensor product of decoupled theories whenever the spacetime field configuration factorizes in an appropriate sense. Besides the dilaton and metric discussed in the introduction, string fields that behave in this way include:
\begin{enumerate}
\item 
The $B$-field, for which the world-sheet action (in the bosonic case) is 
\begin{equation}\label{BfieldWSaction}
S^B[X] = \frac1{2\pi\alpha'}\int d^2\!z\,B_{\mu\nu}(X)\partial X^\mu\bar\partial X^\nu\,.
\end{equation}
If, as a two-form, $B = B_{(1)}+B_{(2)}$, where $B_{(1)}$ lives on $M_{(1)}$ and $B_{(2)}$ on $M_{(2)}$, then clearly $S^B[X^1,X^2]=S^{B_{(1)}}[X^1]+S^{B_{(2)}}[X^2]$. The same applies to the supersymmetric world-sheet.
\item 
The bulk closed-string tachyon of either the bosonic or the type 0 theory. In the bosonic case, the tachyon appears as a potential on the world-sheet, and in the type 0 case as a superpotential:
\begin{align}
S^\phi[X] &= \frac1{2\pi\alpha'}\int d^2\!z\,\phi(X) \qquad \text{(bosonic),}
\\
S^\phi[{\bf X}] &= \frac1{2\pi\alpha'}\int d^2\!zd^2\!\theta\,\phi({\bf X}) \qquad \text{(type 0).}
\end{align}
In the type 0 case $\theta$ and $\tilde\theta$ are the fermionic world-sheet coordinates and ${\bf X^\mu} = X^\mu + i\theta\psi^\mu + i\tilde\theta\tilde\psi^\mu + \theta\tilde\theta F^\mu$ is the supercoordinate. In either case, if $\phi$ splits as in \eqref{phifactorize}, the world-sheet action splits.
\item 
Open-string gauge fields in either the bosonic or superstring theories. These are boundary couplings; for a single D-brane the world-sheet actions are:
\begin{align}
S^A[X] &= \frac1{2\pi\alpha'}\int d\tau A_\mu(X)\partial_\tau X^\mu \qquad \text{(bosonic),} \\
S^A[{\bf X}] &= \frac1{2\pi\alpha'}\int d\tau d\theta\,A_\mu({\bf X})D{\bf X}^\mu \qquad \text{(superstring),}
\end{align}
where $\tau$ is the bosonic boundary coordinate. In the superstring case $\theta$ is the fermionic boundary coordinate, and $D = \partial_\theta + \theta\partial_\tau$. Factorization is clear. For multiple D-branes, the action is slightly more complicated, but factorization occurs as long as $A_1$ and $A_2$ (the $x^1$ and $x^2$ components of $A_\mu$ respectively) commute. The same holds also for the heterotic gauge field.
\item
The bosonic open-string tachyon. This acts as a boundary potential:
\begin{equation}
S^\phi[X] = \frac1{2\pi\alpha'}\int d\tau\,\phi(X)\,.
\end{equation}
\end{enumerate}

Perhaps it is also useful to list a few common string fields to which factorization does \emph{not} apply. Usually this occurs because, even when the spacetime field configuration splits, on the world-sheet it couples both sets of fields $X^1$ and $X^2$ to some third set of fields, and therefore indirectly to each other.
\begin{enumerate}
\item
The open-string tachyon in superstring theories. In these theories there is an additional degree of freedom living on the boundary, a superfield $\Gamma = \eta + \theta F$, where $\eta$ is fermionic and $F$ is auxiliary. This field couples to $\mathbf X$ via the tachyon \cite{Witten:1998cd,Harvey:2000na}:
\begin{equation}
S^\phi[{\bf X},\Gamma] = \frac1{2\pi\alpha'}\int d\tau d\theta\,\phi({\bf X})\Gamma\,.
\end{equation}
Even with $\phi(x^1,x^2) = \phi_{(1)}(x^1) + \phi_{(2)}(x^2)$, both superfields ${\bf X}^1$ and ${\bf X}^2$ will couple to $\Gamma$, and therefore to each other.
\item
Compactification moduli and localized tachyons. For example, let the spacetime field $h(x)$ be the radius-squared of an $S^1$ with coordinate $y$. The world-sheet action will contain the term
\begin{equation}
S^h[X,Y] = \frac1{2\pi\alpha'}\int d^2\!z\,h(X)\partial Y\bar\partial Y\,.
\end{equation}
Again, even if $h(x^1,x^2) = h_{(1)}(x^1) + h_{(2)}(x^2)$, both $X^1$ and $X^2$ will couple to $Y$, and therefore to each other.
\item
The transverse scalars on a D-brane. These are not couplings in the usual sense (i.e.\ coefficients for terms in the world-sheet action), but rather are boundary conditions for the world-sheet scalars. Nothing particular happens to the world-sheet theory when a transverse scalar is a sum of a function of $x^1$ and a function of $x^2$.
\end{enumerate}
Indeed, in those cases where the spacetime action is known (such as the Nambu-Goto action for the transverse scalars), it is easy to check that the equations of motion do not factorize.

Finally, let us mention one technical point concerning the statement that the beta functions for a product of two decoupled quantum field theories factorize. Although this may seem like an utterly obvious statement, its validity actually depends on the renormalization scheme employed. A sufficiently crude regulator (such as enforcing a minimum separation in position space between colliding operators) will yield an apparent coupling in the infrared between two theories that are decoupled in the ultraviolet, in the sense that mixed operators will aquire non-zero coefficients in the effective action. (Such a coupling will be fictitious in the sense that mixed correlators will still vanish.) The point is that a change of renormalization scheme is equivalent to a redefinition of coupling constants, which from the spacetime point of view is a field redefinition, and the property of factorization is not invariant under non-linear field redefinitions. For example, gravity with a dilaton and gravity with a minimally coupled massless scalar field are equivalent theories under a non-linear field redefinition (a change of frame), despite the fact that one theory obeys factorization while the other does not. In what follows, we will assume a scheme (such as one employing dimensional regularization) in which decoupled theories stay decoupled under RG flow, and we will permit ourselves only linear field redefinitions. (There is one exception: shifting a scalar field by an amount proportional to $D$ respects factorization, since $D$ splits the same way a scalar does.)

\subsection{Spacetime action}

In the introduction, we reminded the reader that the one-loop beta functions \eqref{betaG} and \eqref{betaPhi} are derivable from the spacetime action \eqref{NSaction}. Specifically, the beta functions are given by a gradient-flow type relation on the space of couplings (spacetime field configurations):
\begin{equation}\label{gradflow}
\beta^I = -\G^{IJ}\frac{\delta \Sb}{\delta\varphi^J}\,.
\end{equation}
Here the index $I$ stands for the field ($G_{\mu\nu}$ or $\Phi$) as well as the spacetime point $x$. $\G_{IJ}$ is a non-degenerate (but not positive-definite) metric on the space of field configurations, which can be written
\begin{equation}\label{GIJdef}
\G_{IJ}\delta\varphi^I\delta\varphi^J =
\int d^D\!x\,G^{1/2}e^{-2\Phi}
\left(
\delta G_{\mu\nu}\delta G^{\mu\nu} - \frac12(4\delta\Phi-{\delta G_\mu}^\mu)^2
\right).
\end{equation}
Note that this metric is not diagonal in the fields, since it mixes the metric and dilaton perturbations. It is, however, diagonal in the spacetime position, i.e.\ does not contain derivatives acting on $\delta G_{\mu\nu}$ etc. Nor does it contain derivatives of the fields themselves, i.e.\ terms such as $R\delta\Phi^2$. We will refer to these properties as ultralocality.

We will assume in this paper that the beta functions for all the fields we consider are derivable from a spacetime action in the sense of \eqref{gradflow}. We will also assume that the (a priori unknown) metric $\G_{IJ}$ is ultralocal. (The last restriction will be slightly relaxed in the last section, for reasons we will explain there.)

\section{Two-derivative actions}

\subsection{Single scalar field}

A free scalar field $\phi$, with action
\begin{equation}
\Sb = -\frac12\int d^D\!x\left(\partial_\mu\phi\partial^\mu\phi + m^2\phi^2\right)
\end{equation}
and equation of motion
\begin{equation}
\nabla^2\phi - m^2\phi = 0\,
\end{equation}
clearly satisfies factorization. From the discussion of $\phi^4$ theory in the introduction, one might conclude that any interaction term is forbidden, since it would introduce a non-linearity into the equation of motion. We will now see that this is not true.

We begin by writing the most general action up to two derivatives:
\begin{equation}
\Sb = -\int d^D\!x\left(\frac12g(\phi)\partial_\mu\phi\partial^\mu\phi + V(\phi)\right).
\end{equation}
We will assume a flat spacetime, but the calculation would go through unchanged on a fixed curved background metric, so long as it factorized as in \eqref{Gfactorize}. A priori the functions $g$ and $V$ may also depend on the spacetime dimensionality $D$. We will assume that $g(\phi)$ does not vanish. The variational derivative is
\begin{equation}\label{vardev}
\frac{\delta \Sb}{\delta\phi} =
g(\phi)\nabla^2\phi +\frac12g'(\phi)\partial_\mu\phi\partial^\mu\phi -V'(\phi)\,.
\end{equation}
When the field configuration splits as in \eqref{phifactorize}, the terms $\nabla^2\phi$ and $\partial_\mu\phi\partial^\mu\phi$ both split:
\begin{equation}\label{split}
\nabla^2\phi = \nabla_{(1)}^2\phi_{(1)} + \nabla_{(2)}^2\phi_{(2)}\,,\qquad
\partial_\mu\phi\partial^\mu\phi = \partial_1\phi_{(1)}\partial^1\phi_{(1)} + \partial_2\phi_{(2)}\partial^2\phi_{(2)}\,.
\end{equation}
To obtain a beta function that splits, we first need to divide the variational derivative \eqref{vardev} by $g(\phi)$ (up to an unimportant overall constant $k$):
\begin{equation}\label{scalarbeta}
k\beta^\phi = -\nabla^2\phi -\frac{g'(\phi)}{2g(\phi)}\partial_\mu\phi\partial^\mu\phi + \frac{V'(\phi)}{g(\phi)}\,.
\end{equation}
In other words, we take the metric on the space of field configurations to be $\G_{IJ}\delta\varphi^I\delta\varphi^J = k\int d^D\!x\,g(\phi)\delta\phi^2$. Second, we require $g'(\phi)/g(\phi)$ to be constant. Two possibilities are allowed: (1) $g(\phi) = g_0$, and (2) $g(\phi) = g_0 e^\phi$. Here $g_0$ is a non-zero constant, and in case (2) we have used the freedom to do linear field redefinitions. Third, we require $V'(\phi)/g(\phi)$ to be a linear combination of $\phi$ and $D$.

In case (1) we obtain the following action,
\begin{equation}\label{freescalar}
\Sb = -g_0\int d^Dx\left(\frac12\partial_\mu\phi\partial^\mu\phi + \frac12m^2\phi^2 + \alpha D\phi + V_0\right),
\end{equation}
and we are back to the free case. Here $m^2$ and $\alpha$ are arbitrary ($D$-independent) constants. The constant $V_0$ has no effect on the dynamics, and therefore may depend arbitrarily on $D$. If $m^2$ is non-zero, then $\alpha$ can be eliminated by shifting $\phi$ by $\alpha D/m^2$.

In case (2) we find the following action:
\begin{equation}\label{intscalar}
\Sb = -g_0\int d^Dx\left[e^\phi\left(\frac12\partial_\mu\phi\partial^\mu\phi + m^2(\phi-1) + \alpha D\right) + V_0\right].
\end{equation}
This is the most general interacting theory that obeys factorization. Again, if $m^2\neq0$, then $\alpha$ can be eliminated by shifting $\phi$. The free action \eqref{freescalar} may be obtained as a limit of \eqref{intscalar} under a rescaling of $\phi$.

Let us now return to string theory. If we consider $\phi$ to be the dilaton, then the free string theory tells us that $m^2=0$. Up to the undetermined constants $\alpha$, $g_0$, and $V_0$, and with the rescaling $\phi=-2\Phi$, \eqref{intscalar} is indeed the dilaton part of the action \eqref{NSaction}. (We will see below that the free parameter $V_0$ is eliminated once we include gravity.)

On the other hand, if we consider $\phi$ to be the bosonic open-string tachyon, then we know that $\alpha=0$, since $\phi=0$ is a classical solution. The resulting potential has precisely the form of the one derived within Boundary String Field Theory \cite{Gerasimov:2000zp,Kutasov:2000qp}:
\begin{equation}
V(\phi) \sim e^\phi(1-\phi)\,.
\end{equation}
This version of string field theory essentially follows the sigma-model approach, treating the tachyon field as a boundary coupling (specifically, a potential) on the  world-sheet. The property of derivability from an action, as in \eqref{gradflow}, has been proven for the beta functions of renormalizable boundary couplings \cite{Kutasov:2000qp,Friedan:2003yc}, so it is not a surprise that this worked. Nonetheless it is remarkable that the form of the potential could be derived without doing a single world-sheet computation. On the other hand, unlike the honest BSFT computation, we cannot determine $g_0$ or $m^2$, and thereby for example verify the Sen conjecture regarding the depth of the tachyon potential.\footnote{One may be tempted to fix $m^2$ and $g_0$ by matching the mass and (say) cubic vertex as computed from \eqref{intscalar} with those obtained in perturbative string theory. This is not correct for the following reason. Because the tachyon has a string-scale mass, the higher-derivative terms in the action, which we have not computed, are just as important for evaluating the mass and interactions as the two-derivative terms we have computed.}

At this point the reader may object that, when dealing with a field with a string-scale mass such as the open-string tachyon, it is inappropriate to work with an action that is truncated at the two-derivative level (or indeed, at any finite number of derivatives). Certainly no solution to the equations of motion derived from this action can be trusted, other than one that is constant in spacetime. Therefore we must emphasize that our arguments have nowhere depended on finding \emph{solutions} to the equations of motion, that is, field configurations on which the beta functions vanish. Such field configurations necessarily involve derivatives of order one in string units. Rather, we consider field configurations with only small derivatives, and expand the beta functions in powers of derivatives; the fact that those beta functions may be far from vanishing does not affect our arguments.

To avoid cluttering the equations, from this point onward we will often omit overall multiplicative and additive constants, such as $g_0$ and $V_0$ above, which have no effect on the equations of motion. We will also omit the undetermined overall constants multiplying the beta functions, such as $k$ in \eqref{scalarbeta}.

A similar argument shows that a complex scalar with a global $U(1)$ symmetry must be free in order to obey factorization. (In superstring theories, the open strings stretching between a D-brane and an anti-D-brane include a complex tachyon field. However, as explained in Section 2, this string field is not subject to the restriction of world-sheet factorization.)

\subsection{Abelian gauge field and neutral scalar}

For simplicity we restrict our considerations to a single abelian gauge field. The case of a gauge field alone is rather uninteresting, since the only gauge-invariant action with up to two derivatives (that makes sense in a general dimension) is the Maxwell action, whose equations of motion obviously do obey factorization. With the idea of applying the result to the bosonic open-string fields, we therefore consider a gauge field in combination with a neutral scalar.

The general action is:
\begin{equation}
\Sb = -\int d^Dx\left(\frac14f(\phi)F_{\mu\nu}F^{\mu\nu} + \frac12g(\phi)\partial_\mu\phi\partial^\mu\phi + V(\phi)\right).
\end{equation}
We assume that $f(\phi)$ and $g(\phi)$ do not vanish. The variational derivatives are:
\begin{align}
\label{vardevgauge}
\frac{\delta \Sb}{\delta\phi} &= 
g(\phi)\nabla^2\phi +\frac12g'(\phi)\partial_\mu\phi\partial^\mu\phi -V'(\phi)-\frac14f'(\phi)F_{\mu\nu}F^{\mu\nu}\,, \\
\label{Avarderiv}
\frac{\delta \Sb}{\delta A_\nu} &= f(\phi)\partial_\mu F^{\mu\nu} + f'(\phi)\partial_\mu\phi F^{\mu\nu}\,.
\end{align}
The field-space metric $\G_{IJ}$ cannot mix $A_\mu$ and $\phi$ fluctuations (the only combination allowed by Lorentz invariance would be $A^\mu\delta A_\mu\delta\phi$, but that is not gauge invariant). So the beta functions must be:
\begin{align}
\beta^\phi &= -\nabla^2\phi -\frac{g'(\phi)}{2g(\phi)}\partial_\mu\phi\partial^\mu\phi + \frac{V'(\phi)}{g(\phi)} + \frac{f'(\phi)}{4g(\phi)}F_{\mu\nu}F^{\mu\nu}\,. \\
\beta^A_\nu &= -\partial_\mu F^{\mu\nu} - \frac{f'(\phi)}{f(\phi)}\partial_\mu\phi F^{\mu\nu}\,.
\end{align}
In order for $\beta^A$ to factorize properly, $f'(\phi)/f(\phi)$ must be constant. One possibility is $f'(\phi)=0$. In this case the gauge field and the scalar are decoupled; the scalar is as in the previous subsection, and the gauge field is a free Maxwell field. The more interesting possibility is $f(\phi) = f_0e^\phi$. Then, in order for the $F_{\mu\nu}F^{\mu\nu}$ term in $\beta^\phi$ to factorize properly, we must have $g(\phi) = g_0e^\phi$. Finally, $V(\phi)$ is as in case (2) above. All in all, we find that the most general action obeying factorization for a gauge field interacting with a scalar is
\begin{equation}\label{scalargauge}
\Sb = -f_0\int d^D\!x\,e^\phi\left(\frac14 F_{\mu\nu}F^{\mu\nu} + \frac12\partial_\mu\phi\partial^\mu\phi + m^2(\phi-1)+\alpha D\right).
\end{equation}
(We have rescaled $A$ to set $f_0=g_0$.) Equation \eqref{scalargauge} agrees with the result found in BSFT for the coupling between the tachyon and the gauge field \cite{Andreev:2000yn}.

\subsection{Multiple scalars}

Having obtained the form of the bosonic open-string tachyon action by applying the factorization constraint, we would like to see what it says about the closed-string tachyon action. The closed-string tachyon necessarily couples to the dilaton and metric, so we need to consider what factorization says about the action for multiple scalars coupled to gravity. We begin by considering multiple scalars without gravity.

We consider scalars $\phi^i$, with action
\begin{equation}
\Sb = -\int d^D\!x\left(\frac12g_{ij}(\phi)\partial_\mu\phi^i\partial^\mu\phi^j + V(\phi)\right).
\end{equation}
A priori the only condition is that the field-space metric $g_{ij}$ be non-degenerate. From the equations of motion, we learn that factorization requires the components of the Christoffel symbols $\Gamma^i_{jk}$ computed from $g_{ij}$ to be constant, and $\partial_iV = g_{ij}({C^j}_k\phi^k + \alpha^jD)$, where ${C^j}_k$ and $\alpha^j$ are constant. The fact that $\Gamma^i_{jk}$ is constant implies that the components of $\Gamma^i_{ij}=\partial_i\ln g^{1/2}$, ${R^i}_{jkl}$ (the Riemann tensor with one index raised) and $R_{ij}$ (the Ricci tensor) are constant.

In two dimensions we have $R_{ij} = \frac12Rg_{ij}$, implying either (1) $g_{ij} = 2R^{-1}R_{ij}$ or (2) $R=0$ (here $R$ is the Ricci scalar derived from the field space metric $g_{ij}$). In case (1) we have $g_{ij}(\phi) = e^{c_k\phi^k}g_{ij}(0)$, where $c_k$ is constant (and $g_{ij}(0)$ can depend on $D$ only through an overall factor). If $c_k=0$ then $V(\phi) = g_{ij}(\frac12{C^i}_k\phi^j\phi^k + \alpha^i\phi^jD) + V_0$. If $c_k\neq0$ then $V(\phi) = e^{c_k\phi^k}(C(c_k\phi^k-1) + \alpha D)+V_0$. In case (2) it follows that $\Gamma^i_{1j}$ and $\Gamma^i_{2j}$ commute with each other as matrices in $i,j$, and we have $g_{ij}(\phi) = {M_i}^{i'}(\phi){M_j}^{j'}(\phi)g_{i'j'}(0)$, where $M(\phi) = \exp(\phi^i\Gamma^\cdot_{i\cdot})$. We were not able to find the general solution for $V(\phi)$ in explicit form.

In more than two dimensions (i.e.\ with more than two scalars) it's not clear whether there is an explicit form for the general metric with constant Christoffel symbol coefficients. We will not pursue this subject further, since it will turn out that the inclusion of gravity will drastically change the story.

\subsection{Pure gravity}

The action is
\begin{equation}
\Sb = \int d^D\!x\,G^{1/2}\left(R-V_0\right),
\end{equation}
where $V_0$ may depend on $D$. The variational derivative is given by
\begin{equation}
G^{-1/2}\frac{\delta \Sb}{\delta G_{\mu\nu}} = -R^{\mu\nu}+\frac12RG^{\mu\nu} -\frac12V_0G^{\mu\nu}\,.
\end{equation}
The term $RG^{\mu\nu}$ would spoil factorization; to remove it we must use the following field-space metric:
\begin{equation}
\G_{IJ}d\varphi^Id\varphi^J = \int d^D\!x\,G^{1/2}\left(\delta G_{\mu\nu}\delta G^{\mu\nu} - \frac12({\delta G_\mu}^\mu)^2\right),
\end{equation}
yielding
\begin{equation}
\beta^G_{\mu\nu} = R_{\mu\nu} - \frac{V_0}{D-2}G_{\mu\nu}\,.
\end{equation}
For this to factorize, $\Lambda\equiv V_0/(D-2)$ must be constant (independent of $D$). So we obtain the action
\begin{equation}
\Sb = \int d^D\!x\,G^{1/2}\left(R-(D-2)\Lambda\right).
\end{equation}

\subsection{Gravity and single scalar}

We start with the action
\begin{equation}
\Sb = \int d^D\!x\,G^{1/2}\left(f(\phi)R - \frac12 G^{\mu\nu}g(\phi)\partial_\mu\phi\partial_\nu\phi - V(\phi)\right).
\end{equation}
The variation with respect to $G_{\mu\nu}$ is
\begin{equation}
G^{-1/2}\frac{\delta \Sb}{\delta G_{\mu\nu}} = -fR_{\mu\nu}+f'\nabla_\mu\partial_\nu\phi + \left(f''+\frac12g\right)\partial_\mu\phi\partial_\nu\phi + \left(\cdots\right)G_{\mu\nu}\,.
\end{equation}
The third term, proportional to $\partial_\mu\phi\partial_\nu\phi$, would spoil factorization, so its coefficient must vanish. Furthermore, for the first and second terms to factorize, $f'/f$ must be constant. One possibility would be for $f$ itself to be constant, but that would imply that $g$ vanishes, making $\phi$ non-dynamical. Therefore (using a linear field redefinition on $\phi$, and up to an overall factor) we have $2f = -g = e^\phi$. The analysis of the potential term proceeds similarly to the case of the scalar without gravity. The result is the following action:
\begin{equation}\label{gravitygauge}
\Sb = \int d^D\!x\,G^{1/2}e^\phi\left(\frac12R + \frac12\partial_\mu\phi\partial^\mu\phi + m^2(\phi-1)+\alpha D\right)
\end{equation}
(note that there is no $V_0$ term). The field-space metric must be of the form
\begin{equation}
\G_{IJ}d\varphi^Id\varphi^J = 2\int d^D\!x\,G^{1/2}e^\phi\left(\delta G_{\mu\nu}\delta G^{\mu\nu} + \eta({\delta G_\mu}^\mu+2\delta\phi)^2\right),
\end{equation}
yielding the following beta functions:
\begin{align}
\beta^G_{\mu\nu} &= R_{\mu\nu} -\nabla_\mu\partial_\nu\phi + m^2G_{\mu\nu} \\
4\eta\beta^\phi &= \nabla^2\phi +\partial_\mu\phi\partial^\mu\phi - 2m^2\phi + (m^2-2\alpha)D 
-(1+2\eta)G^{\mu\nu}\beta^G_{\mu\nu}\,.
\end{align}

Let us now return to string theory. If we consider $\phi$ to be the dilaton, then, using the additional information that it is massless, we obtain the dilaton-metric part of the action \eqref{NSaction}, up to the substitution $\phi=-2\Phi$ and the undetermined constant $\alpha$. (The constant $\eta$, which appears in the beta functions but not in the action, is also undetermined by factorization. It happens to equal $-\frac12$.) It is straightforward to add the $B$-field; the result is that
\begin{equation}\label{NSactionwithB}
\Sb = 
\int d^D\!x\,G^{1/2}e^{-2\Phi}
\left(
\alpha D+R+4\partial_\mu\Phi\partial^\mu\Phi - \frac1{12}H_{\mu\nu\lambda}H^{\mu\nu\lambda}
\right),
\end{equation}
is the only two-derivative action consistent with factorization for a massless scalar, metric, and $B$-field.

On the other hand, if we consider $\phi$ to be the (closed-string) tachyon, then we do not find an action consistent with what we know, namely that flat space with a vanishing tachyon is a classical solution. However, since the tachyon and dilaton can mix, in principle cancelling non-factorizing terms in each other's beta functions, this setup is too constraining. Therefore, in the next subsection we will ask what happens if we include the tachyon and the dilaton at the same time.

\subsection{Gravity and multiple scalars}

With multiple scalars, the general action is
\begin{equation}
\Sb = \int d^D\!x\,G^{1/2}\left(f(\phi)R - \frac12 G^{\mu\nu}g_{ij}(\phi)\partial_\mu\phi^i\partial_\nu\phi^j - V(\phi)\right),
\end{equation}
and the variation with respect to $G_{\mu\nu}$ is
\begin{equation}
-fR_{\mu\nu}+\partial_if\nabla_\mu\partial_\nu\phi^i + \left(\partial_i\partial_jf+\frac12g_{ij}\right)\partial_\mu\phi^i\partial_\nu\phi^j + \left(\cdots\right)G_{\mu\nu}\,.
\end{equation}
Following the same logic as in the previous subsection, we learn that $f = \frac12e^{c_k\phi^k}$ and $g_{ij} = -c_ic_je^{c_k\phi^k}$. Now we have a problem. The kinetic terms in the Lagrangian, namely
\begin{equation}
\frac12e^{c_k\phi^k}\left(R+G^{\mu\nu}\partial_\mu(c_k\phi^k)\partial_\nu(c_k\phi^k)\right),
\end{equation}
involve the scalars only in the linear combination $c_k\phi^k$. The other scalar degrees of freedom are non-dynamical. In sum, \emph{there is no two-derivative action that respects factorization for multiple scalars coupled to gravity.}

This no-go theorem appears to rule out the various effective actions for the tachyon-metric-dilaton system that have been proposed recently \cite{Yang:2005rw,Yang:2005rx,Genenberg:2006de,Bergman:2006pd,Suyama:2006wx,Suyama:2007vh,Brandenberger:2007xu,Swanson:2008dt}. But does it rule out the possibility of having any action at all for the tachyon? The absence of an action would seem to pose problems for the classical consistency of the bosonic and type 0 theories. There are, however, at least two ways around the theorem. First, it is conceivable that the tachyon action does not contain two-derivative terms, but is instead dynamical due to higher-derivative terms. A more likely possibility is simply that the tachyon mixes in an essential way with massive string fields (in particular, non-scalar ones). Such a mixing could eliminate the non-factorizing terms in the tachyon action. It would be interesting to make this proposal concrete, perhaps first in a toy model.

\section{Born-Infeld-type actions}

It is interesting to try to extend the above analysis to actions with more than two derivatives, ideally even with an infinite number. Technically this is difficult, as the number of unknown functions in the action grows rapidly with the number of derivatives. Therefore in this section we will limit ourselves to actions that contain derivatives only in the form of powers of the first derivative, such as the Nambu-Goto and Born-Infeld actions. (As explained in the introduction, however, the Nambu-Goto action itself does not obey factorization, as the transverse scalars on a D-brane do not represent additive world-sheet couplings.) Such an action is applicable for large field gradients but small second and higher derivatives. We will consider two cases: a single scalar field, and a single abelian gauge field.

\subsection{Single scalar field}

The general action takes the form
\begin{equation}\label{uphiaction}
\Sb = \int d^D\!x\, L(u,\phi)\,, \qquad u \equiv \frac12(\partial_\mu\phi)^2\,.
\end{equation}
Its variation is
\begin{equation}\label{uphivar}
\frac{\delta \Sb}{\delta\phi} =
-\pd Lu\nabla^2\phi - \frac{\partial^2L}{\partial u^2}\partial^\mu\phi\partial_\mu u - 2u\frac{\partial^2L}{\partial u\partial\phi} + \pd L\phi\,.
\end{equation}
The beta function can be a linear combination, with constant coefficients, of the independent quantities $\nabla^2\phi$, $\partial^\mu\phi\partial_\mu u$, $u$, $\phi$, and $D$, all of which factorize. The first two of these quantities, $\nabla^2\phi$ and $\partial^\mu\phi\partial_\mu u$, appear only in the first two terms of \eqref{uphivar}; it follows that $\partial L/\partial u$ and $\partial^2L/\partial u^2$ are proportional to each other, hence $\partial L/\partial u = -g(\phi)e^{C u}$ for some $g(\phi)$ and $C$ (where $C$ cannot depend on $D$). If $C=0$ we have $L = -g(\phi)u - V(\phi)$, bringing us back to the case of two derivatives, which was treated in Section 3 above. Therefore we will assume $C\neq0$, yielding $L = -C^{-1}g(\phi)e^{C u} - V(\phi)$. The last two terms in \eqref{uphivar}, divided by $\partial L/\partial u$, must also split:
\begin{equation}
\left(\pd Lu\right)^{-1}\left( - 2u\frac{\partial^2L}{\partial u\partial\phi} + \pd L\phi\right) =
(C^{-1}-2u)\frac{g'(\phi)}{g(\phi)} + e^{-Cu}\frac{V'(\phi)}{g(\phi)}\,.
\end{equation}
Given that $C^{-1}$ doesn't split but $u$ does, $g'(\phi)$ must vanish. And given that $e^{-Cu}$ doesn't split (additively), $V'(\phi)$ must vanish. We are left with an almost unique Lagrangian:
\begin{equation}\label{uact}
\Sb = \int d^Dx\,e^{\pm (\partial_\mu\phi)^2}
\end{equation}
(where we've rescaled $\phi$). Note that the field-space metric $\G_{IJ}$ here involves derivatives of $\phi$: $\G_{IJ}\delta\varphi^I\delta\varphi^J = \int d^D\!x\,e^{\pm (\partial_\mu\phi)^2}\delta\phi^2$.\footnote{In this derivation we implicitly assumed that the beta function contains the term $\nabla^2\phi$. If we dropped this assumption, and also allowed $\G_{IJ}$ to contain differential operators acting on $\delta\phi$, then any action at all would be allowed.}

It's not clear if \eqref{uact} is the action governing any actual string fields. However, since it does not reduce in the limit of small $u$ to the action \eqref{intscalar}, we learn that none of the higher-derivative corrections to the open-string tachyon action are in the form of functions of $\phi$ and $u$; rather they must all involve more than one derivative acting on $\phi$ (or other string fields).

\subsection{Abelian gauge field}

A straightforward calculation shows that the Born-Infeld action,
\begin{equation}\label{BIaction}
\Sb = \int d^D\!x\,\det\!{}^{1/2}\left({\delta^\mu}_\nu + {F^\mu}_\nu\right),
\end{equation}
obeys factorization. It will turn out, however, that factorization does not uniquely specify this particular action. Rather, as we will show, any Lagrangian that obeys factorization must be either the trace or the determinant of some function of ${F^\mu}_\nu$:
\begin{equation}\label{gaugeresult}
\Sb = \int d^D\!x\,\tr f(F)\,, \qquad
\Sb = \int d^D\!x\,\det f(F)\,.
\end{equation}
Here $f$ is any real function of a  real variable, whose action on the matrix ${F^\mu}_\nu$ is defined in the usual way through its power series. (The function $f$ may not depend on $D$.) For example, the Born-Infeld case \eqref{BIaction} is covered by the second equation in \eqref{gaugeresult}, with $f(x) = (1+x)^{1/2}$.

Unlike in the scalar case above, where $u$ was the only Lorentz-invariant quantity involving the gradient of $\phi$, in the case of the gauge field there is an infinite number of them, namely the traces of all powers of ${F^\mu}_\nu$:
\begin{equation}
u_n \equiv \frac1n\tr F^n\,.
\end{equation}
(Note that $u_n$ vanishes for odd $n$.) In any fixed dimension $D$, only a finite number of the $u_n$ are independent. However, for any finite set of the $u_n$, there is a sufficiently large $D$ such that all the members of the set are independent. Since we are looking for an action that applies to all dimensions (possibly with some explicit dependence on $D$), we will treat all the $u_n$ as independent. Thus the Lagrangian is a priori a function of an infinite number of variables:
\begin{equation}
\Sb = \int d^D\!x\,L(D,u_2,u_4,\dots)\,.
\end{equation}
The variation is as follows:
\begin{equation}
\frac{\delta \Sb}{\delta A_\nu} = 
\sum_{n,m=2,4,\dots} \frac{\partial^2L}{\partial u_m\partial u_n}(F^{n-1})^{\mu\nu}\partial_\mu u_m + 
\sum_{n=2,4,\dots}\pd L{u_n}\partial_\mu(F^{n-1})^{\mu\nu}\,.
\end{equation}
The vectors $(F^{n-1})^{\mu\nu}\partial_\mu u_m$ and $\partial_\mu(F^{n-1})^{\mu\nu}$ factorize and are independent (at sufficiently large $D$, in the same sense that the $u_n$ are). Therefore their coefficients must be proportional to each other:
\begin{equation}\label{gaugeconstraints}
\frac{\partial^2L}{\partial u_m\partial u_n} = E_{mn}g(D,u_2,u_4,\dots), \qquad \pd L{u_n} = C_ng(D,u_2,u_4,\dots)
\end{equation}
(where $E_{mn}$ and $C_n$ are $D$-independent). It follows from the second equation of \eqref{gaugeconstraints} that the Lagrangian depends on the $u_n$ only through the following linear combination:
\begin{equation}\label{fdef}
u\equiv\sum_{n=2,4,\dots}C_nu_n = \tr f(F), \qquad f(x) = \sum_{n=2,4,\dots}\frac{C_n}nx^n\,.
\end{equation}
From this definition, $f$ can be any even function that vanishes at the origin. Actually, both restrictions can be dropped, since the odd part of $f$ doesn't contribute to $u$ anyway, and the constant term contributes only a term proportional to $D$. So far we have $L(D,u_2,u_4,\dots) = L(D,u)$. However, from the first equation of \eqref{gaugeconstraints} we learn that $\partial^2L/\partial u^2$ is proportional to $\partial L/\partial u$. As usual, there are two solutions: $L$ is either linear or exponential in $u$. These give rise to the two possibilities in \eqref{gaugeresult} (in the second case, $f$ in \eqref{gaugeresult} is the exponential of $f$ in \eqref{fdef}).

It would be interesting to see if there is any simple condition which, given the class of actions \eqref{gaugeresult} that obey factorization, uniquely picks out the Born-Infeld action for the gauge field on a D-brane. It would also be interesting to investigate what restrictions factorization places on non-abelian generalizations of the Born-Infeld action. However, we will leave these questions to future work.

\acknowledgments

I would like to thank the following individuals for very useful conversations: A. Lawrence, A. Sen, E. Silverstein, D. Tong, B. Wecht, E. Witten, and B. Zwiebach.

This work was partly supported by DOE grant No.\ DE-FG02-92ER40706, by NSF grant PHY 9870115 (while I was at the Stanford Institute of Theoretical Physics), and by a Pappalardo Fellowship (while I was at the MIT Center for Theoretical Physics).

\bibliography{ref}
\bibliographystyle{JHEP}

\end{document}